# Active Admission Control in a P2P Distributed Environment for Capacity Efficient Livestreaming in Mobile Wireless Networks

Andrei Negulescu
Computer Science and Engineering Department
Santa Clara University
Santa Clara, CA 95053 USA
anegulescu@scu.edu

Weijia Shang
Computer Science and Engineering Department
Santa Clara University
Santa Clara, CA 95053 USA
 wshang@scu.edu

*Abstract*— In this study, the *Active Control in an Intelligent and Distributed Environment (ACIDE)* media distribution model solution and algorithms are proposed for livestreaming in capacity efficient mobile wireless networks. The elements of the ACIDE model are a base station and a cluster formed by a number of peers able to establish peer to peer communications. The cluster peers are selected from a group of users interested in livestreaming the same media. The ACIDE model solution minimizes the bandwidth allocated to a cluster of *n* peers such that an uninterrupted media play for all peers is guaranteed. The livestream media is sent to the peers in *packages* and every media package is divided into *n blocks*. The blocks are distributed to the *n* peers of a cluster in two phases, such that the base station bandwidth is utilized during first phase only. The a*llocated bandwidth*, the amount of bandwidth the base station has to allocate to a cluster, is minimized and its lower bound is equal to the bandwidth required for multicasting. In this study, the ACIDE model is used to address the problem of how to find the maximum number of peers *n*, chosen from a group of *N* users, that can be admitted to a cluster knowing the *given allocated bandwidth*, the amount of bandwidth that a base station allocates to a cluster in advance, prior to admitting users. When users become peers of an ACIDE cluster, the *network capacity*, the total number of users who are able to access live media, increases meaning that network resources are used more efficiently. The problem of finding the maximum number of peers *n* is addressed as an optimization problem, with the objective of having the entire given allocated bandwidth used by the peers admitted to the cluster. This problem is NP-complete and a non-optimal solution is proposed for peers' selection such that all admitted peers play media continuously.

## I. Introduction

New livestreaming media distribution services are continually being developed for mobile wireless networks. As a result of the increasing demand for such services in areas with a high density of users, the bandwidth available per user is reduced. The *network capacity*, the total number of users who are able to access the live media, is also reduced.

This study is exploring the problem of finding the maximum number of peers, *n*, that can be admitted to a cluster when knowing the *given allocated bandwidth*, the amount of bandwidth that a base station allocates to a cluster in advance, prior to admitting users. This means that the base station reserves bandwidth before cluster formation. Our solution to this problem uses the method proposed by the *Active Control in an Intelligent and Distributed Environment* (ACIDE) model, to form clusters of peers utilizing *Peer-to-Peer* (P2P) communications [1]. The ACIDE model minimizes the *allocated bandwidth*, the amount of bandwidth the base station has to allocate to a cluster, such that all *n* peers play livestreamed media continuously. The allocated bandwidth lower bound is the *livestream bandwidth*, the bandwidth used by a base station to distribute live media to one peer [1].

In the ACIDE model, a cluster is formed by *n* peers livestreaming identical media from a base station. The livestream media is divided into packages, and each package is divided into *n* blocks such that the base station sends one block to each cluster peer. Then, each peer sends its blocks to the other *n-1* peers in the cluster, which means that only one package is received by one cluster [1]. The model calculates the sizes of these blocks and the amount of bandwidth that should be allocated to each peer of the cluster for downloading one block. The value of the allocated bandwidth to a single cluster is equal to the sum of the bandwidth values allocated to the peers. In order to increase the network capacity, it must be determined what should be the maximum number of peers *n* admitted to a cluster when the given allocated bandwidth and the number of users interested in the same content are known. We propose a non-optimal algorithm to find the maximum number of peers *n*, such that, as the given allocated bandwidth is approaching the allocated bandwidth, a more efficient bandwidth utilization can be achieved if the number of peers *n* is increasing and the livestream bandwidth is decreasing. The ACIDE model solution and algorithms may be used towards the development of distributed systems services and middleware applications in future mobile wireless networks. The model offers a method for the formation of network capacity efficient heterogeneous clusters (peers with next generation WiFi, 6G, and Bluetooth connectivity capabilities) in high-density areas.

This study is organized as follows. In Section II, the ACIDE model and the bandwidth optimization problem are introduced. Section III presents the number of peers optimization problem formulation, the solution and its complexity. The simulation results analysis is given in Section IV. Section V presents the related work. Section VI concludes the study.

## II. THE ACIDE MODEL

This section presents an overview of the ACIDE model. Definitions and assumptions are given. The problem of minimizing bandwidth and its solution are also reviewed.

In the ACIDE model it is assumed that *N* users are in the coverage area of a base station and they request the same media from the base station. Their movement is limited within the base station coverage area. A number of $n \le N$ users may be admitted to a cluster. Inside a cluster, users become *peers* able to establish P2P communications. The livestream media is sent to the peers as packages. Cluster formation assumes the following three properties [1]. First, all the peers have the *interest property*, such that they livestream identical media from the base station. Second, the peers have the *proximity property*, which means, they are able to establish direct radio communications with each other and with the base station. Third, peers have the *resource property*, meaning that each peer uses two half-duplex radio interfaces, one for download and the other for upload. The media transfer from the base station and other peers is done over the download interface while sending media to other peers in the cluster uses the upload interface. For any-to-any peer connectivity the half duplex interfaces can be reconfigured. It is assumed that the upload bandwidth is less than or equal to the download bandwidth in order to reduce the mobile devices energy consumption. The *livestream bandwidth* is defined as the ratio $\frac{S}{T}$, where *S* is the *media package size* and *T* is the *delay bound*. *T* is defined as the maximum time interval required for the distribution of the media package to a peer such that the peer can play the media continuously. The allocated bandwidth to a cluster would be $n\frac{S}{T}$ if *n* copies of the package are sent to all *n* peers. The ACIDE model assumptions have been introduced as follows [1]:

**Assumption 1**: A cluster is formed by *n* peers. For a peer *i*, with the download and upload bandwidth of $d_i$ and $u_i$ respectively, we assume $u_i \le d_i$, $i = 1,...,n$.

**Assumption 2**: It is assumed that the livestream bandwidth satisfies the following inequality: $\frac{S}{T} < \sum_{i=1}^{n} d_i$.

**Assumption 3**: It is assumed that $u_i \le d_j$ is satisfied for $i = 1,...,n$, $j = 1,...,n$.

Two phases are necessary for the distribution of a package to the peers of a cluster. In Phase 1, a media package is divided into *n* blocks and *bandwidth* $bw_i$ is allocated by the base station to send block *i* to peer *i*. $T_1$ is the package distribution time in Phase 1. During Phase 2, which starts after every peer receives its block from the base station, peer *i* sends block *i* to peer *j* and also receives block *j*, for $j = 1,...,i-1,i+1,...,n$, from the other *n-1* peers. Each peer receives *n-1* blocks from the other peers in *n-1* steps. In each step, a peer establishes two P2P concurrent sessions: one for upload and one for download. Phase 2 sessions are established over direct P2P connections. $T_2$ is the package distribution time in Phase 2. The bandwidth optimization problem has been formulated as [1]:

**Problem 1**: Given a cluster of *n* peers with $d_i$ and $u_i$, the download and upload bandwidth values of peers $i = 1,...,n$, and a known delay bound *T*, how should a package be divided into *n* blocks with sizes $s_1,...,s_i,...,s_n$ and what should be the bandwidth values $bw_i$, allocated to each peer such that the allocated bandwidth $\sum_{i=1}^{n} bw_i$ is minimized? Problem 1 is stated as follows:

$$\text{Minimize} \quad bw = \sum_{i=1}^{n} bw_i$$

$$\text{Subject to} \quad T_1 + T_2 \le T$$

$T_1$ and $T_2$ depend on parameters $d_i$, $u_i$, $s_i$, and $bw_i$. In Phase 1, block *i* is sent by the base station to user *i* with an allocated bandwidth equal to $bw_i$. It is assumed that block *i* is received by user *i* within a delay bound equal to $\frac{s_i}{bw_i}$. Phase 2 communications do not utilize base station bandwidth and start after every cluster peer *i* receives block *i* in Phase 1, therefore:

$$T_1 = \max\{\frac{s_i}{bw_i}, i = 1,....,n\} \qquad (1)$$

It has been proven that the allocated bandwidth is optimal if the times required to transfer *n* blocks to *n* peers in Phase 1 are equal [1]. This result has been introduced as:

**Theorem 1:** The objective function $\sum_{i=1}^{n} bw_i$ is minimized if and only if all events in Phase 1 take the same time and all the events take the same time in Phase 2, during each step.

The optimal solution for $s_i$ and $bw_i$ has been calculated using Theorem 1. Therefore, in Phase 1, we have:

$$bw = \frac{bw_i}{s_i} S, \ i = 1,...,n \quad (2)$$

The following system of $n$ linear equations was derived:

$$\frac{\sum_{i=1}^{k} u_i}{u_k} s_k + \sum_{i=k+1}^{n} s_i = S, \ k = 1,...,n. \quad (3)$$

Let $\alpha_k = \frac{1}{u_k} \sum_{i=1}^{k} u_i$, $k = 2,...,n$, and $\alpha_1 = 0$ for $k = 1$.

Using a matrix notation and the definition of coefficients $\alpha_k$, the system of equations in (3) is represented as:

$$\begin{bmatrix} 1 & 1 & ... & 1 & 1 \\ 0 & \alpha_2 & ... & 1 & 1 \\ ... & ... & ... & ... & ... \\ 0 & 0 & ... & \alpha_{n-1} & 1 \\ 0 & 0 & ... & 0 & \alpha_n \end{bmatrix} \cdot \begin{bmatrix} s_1 \\ s_2 \\ ... \\ s_{n-1} \\ s_n \end{bmatrix} = \begin{bmatrix} S \\ S \\ S \\ S \\ S \end{bmatrix} \quad (4)$$

The optimal sizes $s_i$, $i = 1,...,n$ are calculated from (4). Then the optimal bandwidth derived in [1] is:

$$bw = \sum_{i=1}^{n} bw_i \geq \frac{S}{T - (n-1)\frac{S}{\sum_{i=1}^{n} u_i}} = \frac{S \sum_{i=1}^{n} u_i}{T \cdot \sum_{i=1}^{n} u_i - (n-1)S} \quad (5)$$

In the ACIDE P2P communication model the minimum allocated bandwidth $bw$ is given by (5). One observation is that for an arbitrarily large $n$, the allocated bandwidth is approaching its lower bound $\frac{S}{T}$, the livestream bandwidth, which is also equal to the multicast bandwidth required to send only one package of size $S$ to the entire cluster. From (5), the following result has been proven [1]:

**Corollary 1**: If the bandwidth is minimum, then $\frac{S}{T} \leq \frac{\sum_{i=1}^{n} u_i}{n}$.

Let $\mathbf{s} = [s_1,...,s_n]^t$ and $\mathbf{bw} = [bw_1,...,bw_n]^t$ be the block size and peer allocated bandwidth solution vectors respectively. Then, the solution to Problem 1 is given by the output of the algorithm *MIN_BANDWIDTH* as follows.

**MIN_BANDWIDTH** ($n, T, d_i, u_i$)

1  Create list $L$ with $n$ users
2  Sort $L$ in increasing order of $u_i$, $i = 1,...,n$
3  Calculate the coefficients $\alpha_k = \frac{1}{u_k} \sum_{i=1}^{k} u_i$, $k = 2,...,n$
4  Calculate $\mathbf{s} = [s_1,...,s_n]^t$ using the linear system in (4)
5  Calculate the allocated bandwidth $bw$ for $L$ using (5)
6  Calculate $\mathbf{bw} = [bw_1,...,bw_n]^t$ according to (2)
7  **return** ($bw, T_1, T_2, [s_1,...,s_n]^t, [bw_1,...,bw_n]^t$)

The complexity of the *MIN_BANDWIDTH* algorithm is evaluated next. The matrix in (4) is triangular, which means the time complexity of calculating the values of vector $\mathbf{s} = [s_1,...,s_n]^t$ is $\theta(n^2)$. Because the time complexity of calculating $bw_i$ is $\theta(n)$, we conclude that the overall *MIN_BANDWIDTH* algorithm time complexity is $\theta(n^2)$.

### III. PROBLEM FORMULATION

In this section, the problem of finding the maximum number of peers $n$ that can be admitted to a cluster knowing the given allocated bandwidth $BW$ is formulated as an optimization problem and a solution is proposed. All peers admitted to an ACIDE cluster play livestream media continuously.

Let $N$ be the number of users requesting livestream media from a base station. It is assumed that all $N$ users satisfy the interest, proximity and resource properties introduced in Section II. A cluster of $n \leq N$ peers can be formed according to the ACIDE media distribution model and Theorem 1. The problem of finding the maximum number of users that can be admitted as cluster peers is formulated below.

**Problem 2**: Knowing the given allocated bandwidth $BW$, the number of users $N$ with download and upload bandwidth $d_j$ and $u_j$, $j = 1,...,N$, and the delay bound $T$, what is the maximum number of users $n \leq N$ that can be admitted as cluster peers, how should a package be divided into $n$ blocks with sizes $s_1,...,s_i,...,s_n$ and what is the bandwidth value $bw_i$, $i = 1,...,n$, allocated to each peer such that $\sum_{i=1}^{n} bw_i$ is minimized and $\sum_{i=1}^{n} bw_i \leq BW$ ? In other words, for given parameters $N$, $d_j$ and $u_j$, $j = 1,...,N$, $T$ and $S = \sum_{i=1}^{n} s_i$, how should the given allocated bandwidth $BW$ be divided and allocated to a number of $n \leq N$ peers of an ACIDE cluster and what values should be selected for $s_i$ and $bw_i$, $i = 1,...,n$, such that $n$ is maximized and a continuous media playback on all peers is guaranteed for a minimum $\sum_{i=1}^{n} bw_i \leq BW$ ? Problem 2 is stated as follows:

    Maximize      $n$

    Subject to     $T_1 + T_2 \leq T$

$$bw = \sum_{i=1}^{n} bw_i \leq BW$$

$$n \leq N$$

Dividing the given allocated bandwidth $BW$ among a number of peers $n \leq N$ in Problem 2 is similar to the known *SUBSET_SUM* problem, which means finding the maximum number of users that can be admitted to a cluster such that the minimum allocated bandwidth $bw = \sum_{i=1}^{n} bw_i \leq BW$. Since the *SUBSET_SUM* problem is known to be NP-complete [2], we propose a non-optimal solution for peers' selection and for calculating the maximum number of peers that can be admitted to the cluster. We use the solution to Problem 1 for the following considerations.

From (5), a cluster of $n$ peers allocated bandwidth becomes

$$bw = \sum_{i=1}^{n} bw_i \geq \frac{\frac{S}{T}}{1 - \frac{(n-1)}{\sum_{i=1}^{n} u_i} \frac{S}{T}} \quad \text{and we have} \quad 1 - \frac{(n-1)}{\sum_{i=1}^{n} u_i} \frac{S}{T} \geq \frac{S}{T} \frac{1}{bw}.$$

Then, $\frac{(n-1)}{\sum_{i=1}^{n} u_i} \frac{S}{T} \leq 1 - \frac{S}{T} \frac{1}{bw}$. Therefore, the upper bound of the number of admitted peers $n \leq N$, for an allocated bandwidth $bw = \sum_{i=1}^{n} bw_i \leq BW$, where $i = 1, ..., n$ and $j = 1, ..., N$, depends on the livestream bandwidth and is given by:

$$n \leq 1 + \frac{bw - \frac{S}{T}}{bw \frac{S}{T}} \sum_{j=1}^{N} u_j = 1 + \frac{\sum_{j=1}^{N} u_j}{\frac{S}{T}} - \frac{\sum_{j=1}^{N} u_j}{bw} \quad (6)$$

Because $\frac{S}{T} \leq bw \leq BW$, the lower bound of the given allocated bandwidth $BW$ is equal to the livestream bandwidth $\frac{S}{T}$. From (6), $n = 1$ if $BW = bw = \frac{S}{T}$, meaning that only one peer is admitted to the cluster and livestreaming follows a unicast model. Another observation from (6) is that as the livestream bandwidth $\frac{S}{T}$ increases, for a constant $BW \geq \frac{S}{T}$, the number of admitted peers decreases.

Since $bw$ is minimum, the upper bound value of the livestream bandwidth is calculated according to Corollary 1. Therefore, when $\frac{S}{T}$ reaches its upper bound $\frac{\sum_{j=1}^{N} u_j}{N}$, all $n = N$ peers are admitted if the given allocated bandwidth is $BW = bw = N \frac{S}{T}$.

In order to find the maximum number of admitted peers $n$, we propose *JOIN_CLUSTER*, a non-optimal algorithm that calculates the number of peers optimization problem solution in polynomial time, in several iterations. The algorithm creates a sorted list with $N$ users, in the increasing order of the upload bandwidth $u_j$, $j = 1, ..., N$. The initial value of the minimum allocated bandwidth $bw$ to a cluster formed with all $N$ users is calculated. If $bw > BW$ the user with the lowest upload bandwidth is removed and $bw$ is recalculated for a cluster formed with the remaining users on the list. These actions are repeated for several iterations, until $bw = \sum_{i=1}^{n} bw_i \leq BW$, where $n \leq N$. Using the solution to Problem 1, the minimum allocated bandwidth $bw$ is calculated according to (4), (5) and (2).

The maximum number of iterations of the *JOIN_CLUSTER* algorithm is $N$, meaning that at most $N$ systems of linear equations, as presented in (4), have to be solved. Because the complexity of calculating the allocated bandwidth $bw$ is $\theta(n^2)$ and $1 \leq n \leq N$, the solution to Problem 2 can be calculated with the overall time complexity of $\theta(n^3)$. The algorithm is presented as follows.

**JOIN_CLUSTER** ( $N, BW, d_j, u_j$ )

1. Create list $L$ with $N$ users
2. Sort $L$ in increasing order of $u_j$, $j = 1, ..., N$
3. Initialize $n = N$
4. Calculate the allocated bandwidth $bw = \sum_{i=1}^{n} bw_i$ for $L$
5. **while** ( $bw > BW$ )
6.    Remove first user from $L$
7.    $n = n - 1$
8.    Calculate the allocated bandwidth $bw = \sum_{i=1}^{n} bw_i$ for $L$
9. **endwhile**
10. **return** ( $n, bw, L,$ )

The algorithm calculates the maximum number of peers $n$ and returns $L$, the list of users admitted as cluster peers such that the allocated bandwidth $bw$ is minimized and $\sum_{i=1}^{n} bw_i \leq BW$. In section IV we present the simulation results and evaluate the performance of the *JOIN_CLUSTER* algorithm.

## IV. SIMULATION AND PERFORMANCE EVALUATION

This section presents the performance evaluation of the *JOIN_CLUSTER* peer admission algorithm. The simulation uses the set of parameters and values introduced in [1] for the analysis of the *MIN_BANDWIDTH* algorithm. Two livestream bandwidth values were selected for this study. A set of randomly chosen given allocated bandwidth values has been used. The maximum number of users admitted to a cluster has been calculated for both livestream bandwidth values and each given allocated bandwidth. GNU Octave, a set of tools designed for solving linear algebra problems, was used for simulation.

### A. Simulation Setup

We assumed a delay bound of $T$ = 200ms and cluster sizes of $n \in \{5, 10, 15, 20, 40, 60, 80, 100, 120\}$ peers. Table I presents the bandwidth upload and download ranges for all clusters sizes. The peer bandwidth upload and download values were chosen at random, from the given ranges, considering the following assumption: $u_1 \leq ... \leq u_n \leq \min\{d_1, ..., d_n\}$. The livestream bandwidth values of $\{10000, 12000, 14000, 16000\}$ bps were considered for simulation such that the result of Corollary 1 is satisfied for $n$ = 5 and range $U5$. In our performance evaluation we present the simulator results for the livestream bandwidth values of 10000 bps and 16000 bps.

The optimal block sizes vector solution $\mathbf{s} = [s_1, ..., s_n]^t$ is calculated by our ACIDE simulator. This solution is used to calculate $bw$, $T_1$, $T_2$ and the optimal values of the allocated bandwidth vector, $\mathbf{bw} = [bw_1, ..., bw_n]^t$.

TABLE I. THE UPLOAD AND DOWNLOAD BANDWIDTH RANGES [1]

| Cluster Size $n$ | Upload Bandwidth Range [bps] | Download Bandwidth Range [bps] |
|---|---|---|
| 5 | $U5 = [10000, 20000]$ | $D5 = [20000, 30000]$ |
| 10 | $U10 = [10000, 30000]$ | $D10 = [30000, 50000]$ |
| 15 | $U15 = [10000, 40000]$ | $D15 = [40000, 70000]$ |
| 20 | $U20 = [10000, 50000]$ | $D20 = [50000, 90000]$ |
| 40 | $U40 = [10000, 60000]$ | $D40 = [60000, 110000]$ |
| 60 | $U60 = [10000, 70000]$ | $D60 = [70000, 130000]$ |
| 80 | $U80 = [10000, 80000]$ | $D80 = [80000, 150000]$ |
| 100 | $U100 = [10000, 90000]$ | $D100 = [90000, 170000]$ |
| 120 | $U120 = [10000, 100000]$ | $D120 = [100000, 190000]$ |

### B. Simulation Results

The wireline model presented in [3], [4] has been evaluated and applied to calculate the allocated bandwidth for cluster peers livestreaming in mobile wireless networks. The peers of the wireline model use full-duplex interfaces with upload and download bandwidth values from Table I. We assumed that all peers download media from a base station with the same livestream bandwidth. The objective of this model is to minimize the package distribution time from the base station to the peers. Fig. 1 presents the variation of the media package distribution time values with the number of peers of a cluster for the wireline model. The minimum distribution times of clusters with sizes of 5, 10, 15 and 20 peers are graphed for four different livestream bandwidth values. As shown in Fig. 1 the package distribution time is reduced as the cluster size is increased, and is getting larger with the livestream bandwidth.

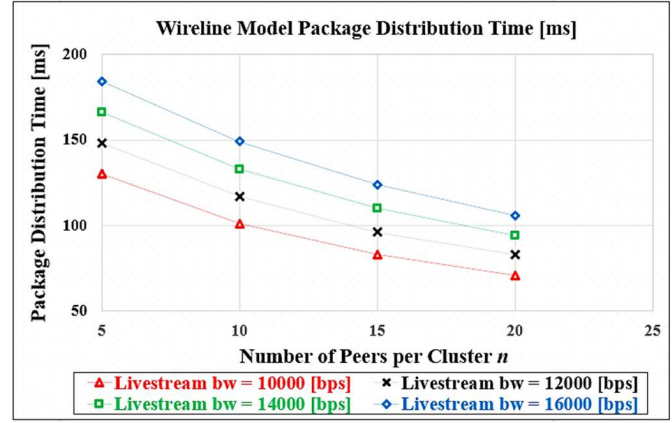

Fig. 1. Distribution time values calculated for the wireline model

In comparison, in our ACIDE model the allocated bandwidth is reduced as the number of peers increases and a constant package distribution time is guaranteed [1]. According to Theorem 1, the events in Phase 1 take the same time. This result is shown in Fig. 2. We observe that the block size to bandwidth ratios $\dfrac{s_i}{bw_i}$ are equal to $T_1$ for all $i = 1, ..., n$ peers of a cluster with size $n$ = 5.

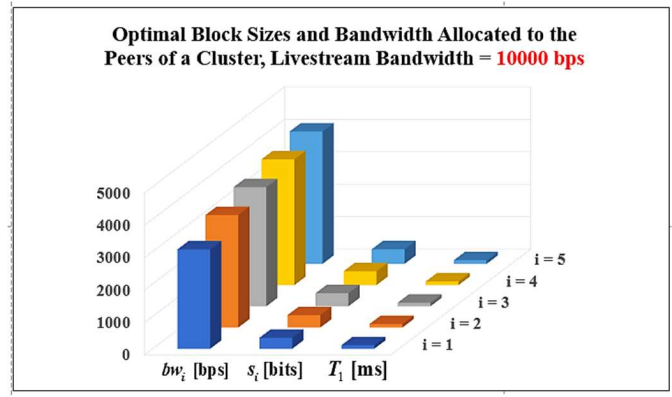

Fig. 2. Block sizes and bandwidth allocated to the peers of a cluster, $n$=5

When the size of a cluster is getting larger, the allocated bandwidth is decreasing and the base station bandwidth is utilized more efficiently, such that for arbitrarily large cluster sizes, the values of the allocated bandwidth are approaching the base station livestream bandwidth. As the livestream bandwidth increases, more bandwidth is allocated to same size clusters. When the livestream bandwidth reaches the Corollary 1 upper bound, the allocated bandwidth becomes equal to $n\dfrac{S}{T}$, which is the bandwidth allocated in a unicast model. This implies that the base station bandwidth is utilized less efficiently as the livestream bandwidth approaches its upper bound.

The admission control mechanism of the ACIDE cluster is using time $T_2$ as the allocated bandwidth control loop variable. $T_2$ is reduced as the number of peers admitted to the cluster is increasing. When $T_2$ is decreasing, in order to satisfy the delay bound $T$, time $T_1$ is increased. From (1) and Theorem 1, the allocated bandwidth is reduced and approaches the livestream bandwidth. Therefore, in our *JOIN_CLUSTER* algorithm, we decided to prioritize removing the peer with the lowest upload bandwidth. While the number of users $n$ admitted to the cluster is not optimal, it is the maximum number of users offering the most efficient utilization of the given allocated bandwidth.

Fig. 3 graphs the optimal sizes of the blocks allocated to the peers of a cluster with size $n$, $20 \leq n \leq 120$. The average rate of change of the sizes of blocks allocated to the peers of a particular cluster is decreasing as the cluster size increases. As the cluster size $n$ and the upload bandwidth $u_i$ increase, the optimal block sizes $s_i$ decrease and less bandwidth $bw_i$ is allocated to the peers of the cluster. In the ACIDE model, the package distribution time is guaranteed to be constant, regardless of the livestream bandwidth value, such that the peers of a cluster play livestream media continuously.

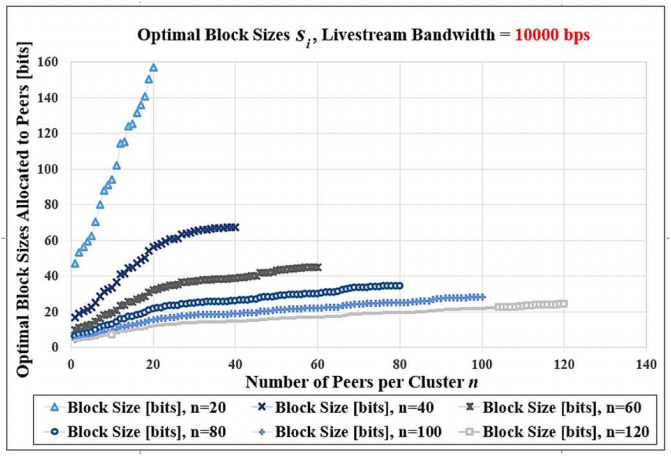

Fig. 3. The optimal block sizes allocated to the peers of an ACIDE cluster

In order to analyze how $n$, the maximum number of users admitted to a cluster, is changing with the given allocated bandwidth $BW$, we have run the *JOIN_CLUSTER* algorithm on our simulator for the livestream bandwidth values of 10000 bps and 16000 bps. The given allocated bandwidth values $BW$ have been chosen equal to the livestream bandwidth values of $\frac{S}{T}$, such that the algorithm runs for $N$ iterations. The variation of the $N$ calculated values, representing the number of peers admitted to an ACIDE cluster, for the two chosen livestream bandwidth values are graphed in Fig. 4. We assumed that the livestreaming requests have been made by users with the upload and download bandwidth ranges from Table I. The maximum cluster sizes are chosen as $N = \{5, 10, 15, 20, 40, 60\}$. In each iteration, the allocated bandwidth has been calculated for a cluster of size $n$, where $1 \leq n \leq N$. In Fig. 4 we observe that for a random, fixed value of $BW$, the number of peers $n$ admitted to the cluster decreases as the livestream bandwidth $\frac{S}{T}$ increases.

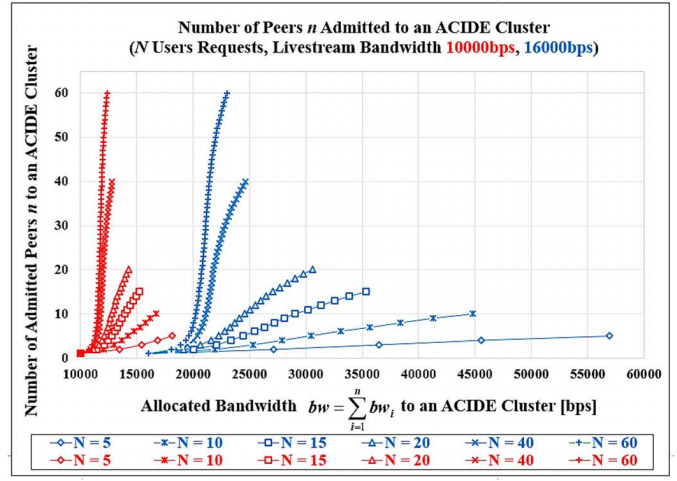

Fig. 4. Number of peers $n$ admitted to an ACIDE cluster for the livestream bandwidth values of 10000 bps and 16000 bps

The rate of change of $n$, the number of admitted peers, decreases when the livestream bandwidth increases. This implies that in order to admit more peers, the given allocated bandwidth should become larger. Therefore, the *bandwidth allocation efficiency*, the percentage of the given allocated bandwidth utilized by a cluster, given by $\frac{bw}{BW}$, is reduced as $\frac{S}{T}$ approaches its upper bound. One observation from Fig. 4 is that as the livestream bandwidth decreases the given allocated bandwidth for which all $N$ peers are admitted can be reduced, meaning that the admission algorithm bandwidth allocation efficiency is increasing as the livestream bandwidth decreases. We have used six given allocated bandwidth values $BW$. They are shown in Table II and Table III and they have been chosen at random for each of the two livestream bandwidth values of 10000bps, 16000bps. Fig. 5 and Fig. 6 graph the maximum number of peers $n$ admitted to a cluster for the six $BW$ values.

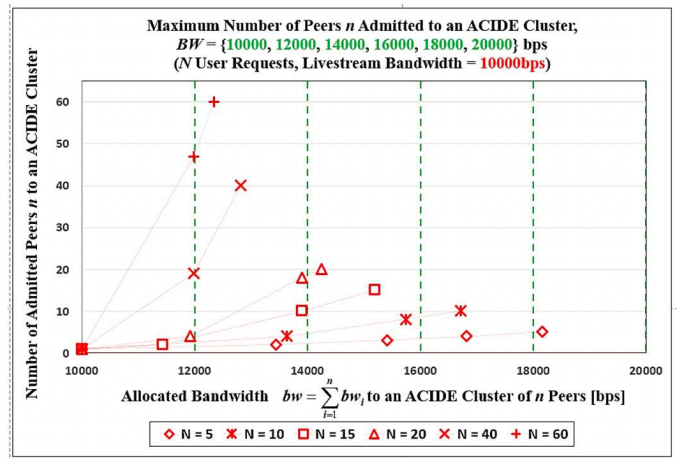

Fig. 5. Maximum number of peers $n$ admitted for different given allocated bandwidth values and a livestream bandwidth of 10000 bps

TABLE II. JOIN_CLUSTER Algorithm: Allocated Bandwidth Performance for A Livestream Bandwidth of 10000 *bps*

| N User Requests (Livestream bandwidth = 10000 [bps]) | Given Allocated Bandwidth BW 20000 [bps] | | Given Allocated Bandwidth BW 18000 [bps] | | Given Allocated Bandwidth BW 16000 [bps] | | Given Allocated Bandwidth BW 14000 [bps] | | Given Allocated Bandwidth BW 12000 [bps] | | Given Allocated Bandwidth BW 10000 [bps] | |
|---|---|---|---|---|---|---|---|---|---|---|---|---|
| | $n$ | $\frac{bw}{BW}$ [%] | $n$ | $\frac{bw}{BW}$ [%] | $n$ | $\frac{bw}{BW}$ [%] | $n$ | $\frac{bw}{BW}$ [%] | $n$ | $\frac{bw}{BW}$ [%] | $n$ | $\frac{bw}{BW}$ [%] |
| 5 | 5 | 90.81 | 4 | 93.43 | 3 | 96.28 | 2 | 96.05 | 1 | 83.33 | 1 | 100 |
| 10 | 10 | 83.58 | 10 | 92.86 | 8 | 98.36 | 4 | 97.40 | 1 | 83.33 | 1 | 100 |
| 15 | 15 | 75.98 | 15 | 84.42 | 15 | 94.97 | 10 | 99.33 | 2 | 95.30 | 1 | 100 |
| 20 | 20 | 71.23 | 20 | 79.14 | 20 | 89.04 | 18 | 99.31 | 4 | 99.34 | 1 | 100 |
| 40 | 40 | 64.05 | 40 | 71.17 | 40 | 80.07 | 40 | 91.50 | 19 | 99.83 | 1 | 100 |
| 60 | 60 | 61.74 | 60 | 68.60 | 60 | 77.17 | 60 | 88.20 | 47 | 99.91 | 1 | 100 |

TABLE III. JOIN_CLUSTER Algorithm: Allocated Bandwidth Performance for A Livestream Bandwidth of 16000 *bps*

| N User Requests (Livestream bandwidth = 16000 [bps]) | Given Allocated Bandwidth BW 60000 [bps] | | Given Allocated Bandwidth BW 50000 [bps] | | Given Allocated Bandwidth BW 40000 [bps] | | Given Allocated Bandwidth BW 30000 [bps] | | Given Allocated Bandwidth BW 20000 [bps] | | Given Allocated Bandwidth BW 16000 [bps] | |
|---|---|---|---|---|---|---|---|---|---|---|---|---|
| | $n$ | $\frac{bw}{BW}$ [%] | $n$ | $\frac{bw}{BW}$ [%] | $n$ | $\frac{bw}{BW}$ [%] | $n$ | $\frac{bw}{BW}$ [%] | $n$ | $\frac{bw}{BW}$ [%] | $n$ | $\frac{bw}{BW}$ [%] |
| 5 | 5 | 94.93 | 4 | 91.07 | 3 | 91.20 | 2 | 90.43 | 1 | 80.00 | 1 | 100 |
| 10 | 10 | 74.66 | 10 | 89.60 | 8 | 96.00 | 4 | 93.02 | 1 | 80.00 | 1 | 100 |
| 15 | 15 | 58.88 | 15 | 70.66 | 15 | 88.32 | 10 | 96.87 | 1 | 80.00 | 1 | 100 |
| 20 | 20 | 50.98 | 20 | 61.17 | 20 | 76.47 | 19 | 99.26 | 2 | 95.61 | 1 | 100 |
| 40 | 40 | 41.09 | 40 | 49.31 | 40 | 61.64 | 40 | 82.18 | 3 | 97.34 | 1 | 100 |
| 60 | 60 | 38.33 | 60 | 45.99 | 60 | 57.49 | 60 | 76.66 | 7 | 99.70 | 1 | 100 |

We observe that as *N* increases and *BW* is getting reduced, the bandwidth allocation efficiency increases, which means, less bandwidth is necessary for a continuous livestreaming inside the cluster and a larger percentage of the given bandwidth is allocated to the peers. Also, as *N* is getting larger, a small change in *BW* will result in a large decrease in the number of admitted peers. Since the rate of change of *n* is higher as the livestream bandwidth decreases, for smaller variations of *BW* we note larger differences in the number of admitted peers.

The given allocated bandwidth performance measurements are presented in Table II and Table III. We observe that if the given allocated bandwidth $BW = bw = \sum_{i=1}^{N} bw_i$ then all *N* users are admitted as cluster peers and the entire bandwidth is allocated to the *N* peers. For a given livestream bandwidth and constant $BW \geq \frac{S}{T}$ we observe that as *N* increases the allocated bandwidth $bw = \sum_{i=1}^{N} bw_i$ is approaching its lower bound $\frac{S}{T}$. In this case too much bandwidth is pre-allocated, or the given allocated bandwidth is too large, in order to admit all users as peers. This results in a lower bandwidth allocation efficiency.

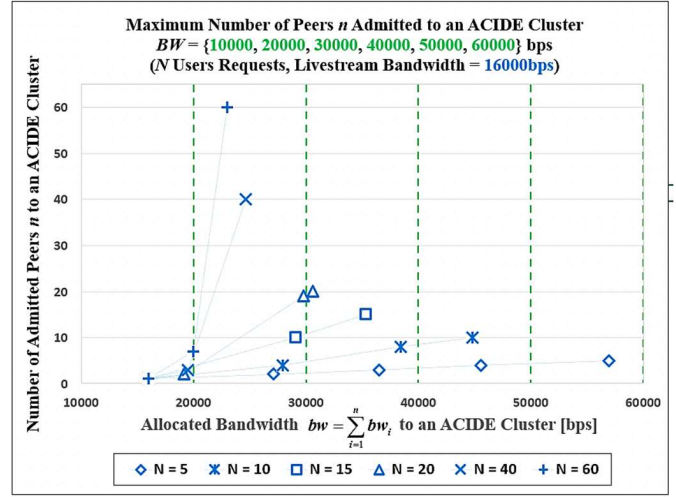

Fig. 6 Maximum number of peers *n* admitted for different given allocated bandwidth values and a livestream bandwidth of 16000 bps

In summary, our *JOIN_CLUSTER* algorithm analysis indicates that as the livestream bandwidth decreases and the number of users *N* is getting larger, a higher percentage of the given allocated bandwidth is utilized by the maximum number of peers *n* admitted to the cluster.

## V. Related Work And Motivation

Motivated by P2P communication systems attributes, such as being collaborative, distributed and scalable, recent research studies on how to build a P2P mobile wireless network capable of efficient media dissemination, have been dedicated to the design of P2P overlay architectures and routing protocols. The bandwidth utilization and its variability have been identified as important challenges and overlay routing solutions have been proposed in [5], [6] to improve transmission performance between nodes. Non-linear algorithms have been analyzed in [7], [8], [9] for node resource optimization in bandwidth variable P2P communications, and node selection algorithms have been proposed for increasing P2P networks' collaboration efficiency [10], [11]. Data distribution performance studies on P2P wireless and wired overlay topologies indicate that the energy consumption, bandwidth utilization and latency increase with the number of participating nodes [12], [13]. Resource allocation, user grouping algorithms and media multicast transmission methods in wireless networks have been presented in [14], [15], [16], [17] in an attempt to reduce the bandwidth utilized by video streaming applications. Better performance results on energy consumption and bandwidth utilization for high user density wireless networks have been reached using D2D solutions, or multicast user grouping algorithms in hybrid and wireless heterogeneous networks [18], [19], [20].

In this study our objective is to maximize the number of peers $n$ that can be admitted to an ACIDE P2P cluster such that the given allocated bandwidth is utilized efficiently, whereas the above work focus is placed on reducing users' bandwidth utilization for media streaming applications. The bandwidth efficient ACIDE model is used to increase the capacity of a wireless network where $N$ peers livestream identical media from a base station [1]. The ACIDE model borrows the concepts of P2P cluster communications and media object segmentation presented in [3], [4]. The peer admission control problem is formulated as an optimization problem. Because the problem is NP-complete we propose a non-optimal algorithm that calculates the maximum number of peers admitted to an ACIDE cluster. Our study also evaluates the bandwidth allocation efficiency variation with the livestream bandwidth.

## VI. Conclusion And Future Work

A growing interest in media livestreaming services creates bandwidth availability challenges and limits the network capacity in high density mobile wireless networks. In this study, we formulate the network capacity optimization problem as how to find the maximum number of peers that can be admitted to a cluster knowing the given allocated bandwidth. This problem has been identified as being NP-complete and a non-optimal solution has been proposed. We have found that when applying our solution, as the livestream bandwidth decreases a larger number of peers are admitted for a given allocated bandwidth, with an improved bandwidth allocation efficiency.

As future work, solutions making the ACIDE model based livestreaming possible in multi-hop mobile wireless networks are being investigated. Cluster formation algorithms and peers' mobility control protocols are also being evaluated.

## References

[1] Negulescu A., Shang W. Bandwidth Efficient Livestreaming in Mobile Wireless Networks: A Peer-to-Peer ACIDE Solution, 2023 Oct 20; Preprint available from https://arxiv.org/abs/2310.14283 [cs.NI]

[2] Cormen, Thomas H., Charles E. Leiserson, Ronald L. Rivest, and Clifford Stein.,2022 *Introduction to algorithms*. MIT press

[3] Cui H, Su X, Shang W. An optimal media distribution algorithm in P2P-based IPTV. In2008 Third International Conference on Communications and Networking in China 2008 Aug 25 (pp. 360-364). IEEE.

[4] Cui H, Su X, Shang W. Optimal dissemination of layered videos in P2P-Based IPTV networks. In2009 IEEE International Conference on Multimedia and Expo 2009 Jun 28 (pp. 738-741). IEEE.

[5] Deokate B, Lal C, Trček D, Conti M. Mobility-aware cross-layer routing for peer-to-peer networks. Computers and Electrical Engineering. 2019 Jan 1;73:209–26.

[6] Toce A, Mowshowitz A, Kawaguchi A, Stone P, Dantressangle P, Bent G. An efficient hypercube labeling schema for dynamic Peer-to-Peer networks. J Parallel Distrib Comput. 2017 Apr 1;102:186–98.

[7] Bof N, Carli R, Notarstefano G, Schenato L, Varagnolo D. Newton-Raphson Consensus under asynchronous and lossy communications for peer-to-peer networks. 2017 Jul 28; Available from: http://arxiv.org/abs/1707.09178

[8] Asghari S, Navimipour NJ. Resource discovery in the peer to peer networks using an inverted ant colony optimization algorithm. Peer Peer Netw Appl. 2019 Jan 1;12(1):129–42.

[9] Aslani R, Hakami V, Dehghan M. A token-based incentive mechanism for video streaming applications in peer-to-peer networks. Multimed Tools Appl. 2018 Jun 1;77(12):14625–53.

[10] Disterhoft A, Graffi K. Capsearch: capacity-based search in highly dynamic peer-to-peer networks. In: Proceedings - International Conference on Advanced Information Networking and Applications, AINA. Institute of Electrical and Electronics Engineers Inc.; 2017. p. 621–30.

[11] Qu D, Wu S, Liang D, Zheng J, Kang L, Shen H. Node Cooperation Analysis in Mobile Peer-to-peer Networks. InICC 2020-2020 IEEE International Conference on Communications (ICC) 2020 Jun 7 (pp. 1-6). IEEE.

[12] Pan MS, Lin YP. Efficient data dissemination for Wi-Fi peer-to-peer networks by unicasting among Wi-Fi P2P groups. Wireless Networks. 2018 Nov 1;24(8):3063–81.

[13] Fortuna R, Leonardi E, Mellia M, Meo M, Traverso S. QoE in pull based P2P-TV systems: Overlay topology design tradeoffs. In: 2010 IEEE 10th International Conference on Peer-to-Peer Computing, P2P 2010 – Proceedings. 2010.

[14] Chen S, Yang B, Yang J, Hanzo L. Dynamic Resource Allocation for Scalable Video Multirate Multicast over Wireless Networks. IEEE Trans Veh Technol. 2020 Sep 1;69(9):10227–41.

[15] Yuan Y, Zhang Z, Liu D. AG-MS: A user grouping scheme for DASH multicast over wireless networks. In2017 IEEE 85th Vehicular Technology Conference (VTC Spring) 2017 Jun 4 (pp. 1-5). IEEE.

[16] Guo J, Gong X, Liang J, Wang W, Que X. An Optimized Hybrid Unicast/Multicast Adaptive Video Streaming Scheme over MBMS-Enabled Wireless Networks. IEEE Transactions on Broadcasting. 2018 Dec 1;64(4):791–802.

[17] Katti S, Rahul H, Hu Dina Katabi W, Crowcroft J. XORs in The Air: Practical Wireless Network Coding. 2006.

[18] Araniti G, Scopelliti P, Muntean GM, Iera A. A Hybrid Unicast-Multicast Network Selection for Video Deliveries in Dense Heterogeneous Network Environments. IEEE Transactions on Broadcasting. 2019 Mar 1;65(1):83–93.

[19] Maheswari BU, Ramesh TK. An Improved Delay-Resistant and Reliable Hybrid Overlay for Peer-to-Peer Video Streaming in Wired and Wireless Networks. IEEE Access. 2018;6:56539–50.

[20] Niu Y, Yu L, Li Y, Zhong Z, Ai B. Device-to-Device Communications Enabled Multicast Scheduling for mmWave Small Cells Using Multi-Level Codebooks. IEEE Trans Veh Technol. 2019 Mar 1;68(3):2724–38.